\documentstyle[epsfig]{jaa}

\DeclareMathAlphabet{\mathsc}{OT1}{cmr}{m}{sc}
\def\testbx{bx}%
\DeclareRobustCommand{\ion}[2]{%
\relax\ifmmode
\ifx\testbx\f@series
{\mathbf{#1\,\mathsc{#2}}}\else
{\mathrm{#1\,\mathsc{#2}}}\fi
\else\textup{#1\,{\mdseries\textsc{#2}}}%
\fi}

\def\ch{\footnotesize}
\def\HI{\ion{H}{i}~}

\def\aa{Astron. Astrophys.}
\def\ApJ{Astrophys. J.}
\def\MNRAS{Mon. Not. R. Astron. Soc.}
\def\AJ{Astron. J.}
\def\ARAA{Ann. Rev. Astron. Astrophys.}

\begin{document}
\title[Eridanus group : Radio continuum \& FIR emission]
{Radio continuum and far-infrared emission from the galaxies in the Eridanus group}
\author[Omar \& Dwarakanath] 
{A. Omar\thanks{Present address : ARIES, Manora peak, Nainital, 263 129, Uttaranchal, India}
\thanks{e-mail: aomar@upso.ernet.in}
\& K.S. Dwarakanath
\thanks{e-mail: dwaraka@rri.res.in}\\
Raman Research Institute, Sadashivanagar, Bangalore 560 080, India \\}

\pubyear{xxxx}
\volume{xx}
\date{Received xxx; accepted xxx}
\maketitle
\label{firstpage}
\begin{abstract}

The Eridanus galaxies follow the well-known radio-FIR correlation.  Majority
(70\%) of these galaxies have their star formation rates below that of the Milky Way.
The galaxies having a significant excess of radio emission are identified as low
luminosity AGNs based on their radio morphologies obtained from the GMRT
observations. There are no powerful AGNs ($L_{20cm} > 10^{23}$~W~Hz$^{-1}$) in
the group. The two most far-infrared and radio luminous galaxies in the group have
optical and \HI morphologies suggestive of recent tidal interactions. The
Eridanus group also has two far-infrared luminous but radio-deficient galaxies.
It is believed that these galaxies are observed within a few Myr of the onset
of an intense star formation episode after being quiescent for at least a 100
Myr. The upper end  of the radio luminosity distribution of the Eridanus
galaxies ($L_{20cm} \sim 10^{22}$~W~Hz$^{-1}$) is consistent with that of the
field galaxies, other groups, and late-type galaxies in nearby clusters.

\end{abstract} 

\section{Introduction}

There is a well-known correlation between the non-thermal radio emission and the
thermal far-infrared (FIR) emission from star forming galaxies (van der Kruit
1973, Condon 1992, Yun, Reddy \& Condon 2001).  The radio-FIR correlation has been
observed to be extremely tight ($\sigma \sim 0.26$ dex) over a wide range
($10^{9}-10^{12}$~L$_{\odot}$) of FIR luminosities (Yun, Reddy \& Condon 2001).  The
origin and the tightness of the radio-FIR correlation is not completely
understood. The radio-FIR correlation is linked to the star formation
activities in galaxies, and is not displayed by Active Galactic Nuclei (AGNs).
According to one scenario, ultra-violet photons from massive stars ($M
>8~M_{\odot}$) heat the surrounding dust which emits in the far-infrared, and the
same population of massive stars undergo supernova explosions  thereby
accelerating cosmic electrons responsible for much of the radio emission from
galaxies (Harwit \& Pacini 1975). The presence or absence of the correlation
therefore can be used to make a distinction between the star formation and the
AGN related radio continuum emission in galaxies. 

The radio-FIR correlation has also been used to study the effects of
environment on the evolution of radio sources in galaxy clusters and groups
(e.g., Scodeggio \& Gavazzi 1993, Gavazzi \& Boselli 1999, Miller \& Owen 2001,
Menon 1995, Reddy \& Yun  2004). Studies indicate that normal star forming
galaxies in the local Universe seldom have their 20~cm spectral luminosities
exceeding $10^{23}$~W~Hz$^{-1}$, but AGNs are observed with higher radio
luminosities (Condon 1989, Yun, Reddy \& Condon 2001). Studies in clusters of galaxies
indicate that the luminous radio AGNs ($L_{20cm} > 10^{23}$~W~Hz$^{-1}$) are
hosted mostly by early type galaxies in the cores of clusters (e.g., Gavazzi \&
Boselli 1999, Reddy \& Yun 2004). These radio-luminous AGNs are more commonly
found in the cluster environment than in the field. These studies indicate that
the upper end of the radio luminosity distribution of late type galaxies does
not differ much in different environments.

The Eridanus group was recently studied in detail (Omar 2004, Omar \&
Dwarakanath 2004a,b). This group has $\sim200$ identified members. This group was
predicted to be in an early phase of the cluster formation where different
sub-groups are merging together (Willmer et al. 1989). Although the entire
group is not dynamically relaxed, the galaxy evolution process seems to be
effective in the group. A significant \HI deficiency up to a factor of $2-3$ was
observed in the Eridanus galaxies in the high galaxy density regions (Omar \&
Dwarakanath 2004b). This group also has a significantly large (30\%) population
of early types (E+S0's). In this paper, the radio-FIR correlation for the
Eridanus galaxies is constructed using  the FIR data from the {\it Infrared
Astronomical Satellite} (IRAS; Neugebauer et al. 1984) survey and the 20~cm radio
data from the {\it Northern VLA Sky Survey} (NVSS; Condon et al. 1998). The higher
resolution ($5''-10''$) radio continuum images at 20~cm from the GMRT
observations are used to study the nature of the galaxies departing from the
radio-FIR correlation. The GMRT radio continuum images of some of the normal star
forming galaxies are also presented.

\section{The FIR and the radio data}

\subsection{The IRAS data}

The IRAS faint source catalog has a limiting sensitivity of $\sim0.3$~Jy at
$\lambda 60\mu$. The $60\mu$ flux density, $S_{60\mu}$ is converted to the
luminosity, $L_{60\mu}$ using the relation:

\begin{equation}
log \frac{L_{60\mu}}{L_{\odot}} = 6.014 + 2log\frac{D}{\mathrm Mpc} +
log\frac{S_{60\mu}}{\mathrm Jy}
\label{equ:FIRLUM}
\end{equation}

At the distance to the Eridanus group ($D\sim23$~Mpc), the limiting sensitivity
of the  $IRAS$ observations is $L_{60\mu} \sim 10^{8}$~L$_{\odot}$.  The total
FIR luminosity, $L_{\mathrm FIR}$ is estimated using the relation (Helou et al.
1998):

\begin{equation}
L_{\mathrm FIR} (L_{\odot}) = \left(1 + \frac{S_{100\mu}}{2.58~S_{60\mu}}
\right) L_{60\mu}
\end{equation}

\subsection{The NVSS data}

The 20~cm flux densities are from the NVSS made at a resolution of $45''$
($\sim 5$~kpc) with an rms of 0.5~mJy~beam$^{-1}$. The 20~cm spectral
luminosity, $L_{20cm}$ is estimated using the relation:

\begin{equation}
log \frac{L_{20cm}}{WHz^{-1}} = 20.08 + 2log\frac{D}{Mpc} + log\frac{S_{20cm}}{Jy}
\label{equ:RADLUM}
\end{equation}

The $5\sigma$ sensitivity to the 1.4~GHz spectral luminosity is $\sim1.6\times
10^{20}$~W~Hz$^{-1}$.

\subsection{The GMRT observations}

A total of 57 galaxies in the Eridanus group were observed  by Omar \&
Dwarakanath (2004a) using the GMRT.  The observing strategy  was optimized to
get uniform distribution of  visibilities. Two galaxies were observed
alternately for $15-20$ minutes each followed by $5-7$ minutes of observations of
the secondary calibrators. This cycle was repeated and a total of 3-4 hour of
observing time was accumulated on each galaxy. Majority of these observations
were carried out over a bandwidth of 8~MHz divided into 128 spectral channels.
These observations were primarily aimed at detecting \HI emission from the
Eridanus galaxies. No pre-selection of galaxies  was made based on their radio
or optical properties. Both early type and late type galaxies were observed.
The data were corrected for the spectral response of the IF filters. The
channels devoid of \HI line signals were averaged to obtain the radio continuum
data. The GMRT has a mix of both short and long baselines which make the data
sensitive to extended emission and at the same time also make it possible to
obtain a high resolution image. The data were analysed following the standard
procedures using the {\it Astronomical Image Processing System} 
({\sf AIPS} ) developed by the National Radio Astronomy Observatory (NRAO). The flux density
scale of these observations is based on the standard VLA calibrators 3C~48 and
3C~147. Images were {\sf CLEANed} and self-calibrated to improve the dynamic
range. The images presented here are at angular resolutions which emphasize the
extended radio continuum emission from the galaxies. The details of the image
properties are given in Tab.~\ref{tab:gmrt}. The total flux densities are
estimated from images at a resolution of $\sim1'$. The GMRT flux densities are
consistent with the NVSS flux densities within 20\% for most of the galaxies
except for the two weaker ones, viz., IC~1953 and ESO~548- G 036 where the
uncertainties are  25\% and 30\% respectively.

\begin{table}
\begin{center}
\caption{GMRT image parameters}
\label{tab:gmrt}
\begin{tabular}{lcccl}
\hline
\hline
\ch {\bf Galaxy} & \ch \bf{rms} & \ch \bf{Resolution} & \ch \bf{S$_{\mathrm total}$} & \ch \bf{Morph.} \\
               & \ch (mJy/bm)  & \ch (arc sec)  &\ch (mJy) &\\
\hline
NGC~1407 & 0.50 & $6 \times 6$    & $99\pm10$       & Diffuse \\
NGC~1371 & 0.22 & $15 \times 15$  & $19.7\pm2$      & Linear \\
NGC~1415 & 0.12 & $4 \times 4$    & $27\pm3$        & Linear \\ 
NGC~1482 & 0.81 & $8 \times 8$    & $280\pm30$      & Diffuse \\
NGC~1385 & 0.41 & $15 \times 15$  & $180\pm20$      & Diffuse \\
NGC~1377 & 0.20 & $15 \times 15$  & $<1~(5\sigma)$  &  -- \\
IC~1953  & 0.30 & $50 \times 45$  & $9\pm2$         & Diffuse \\
NGC~1309 & 0.35 & $15 \times 15$        & $68\pm7$        & Diffuse \\
NGC~1345 & 0.23 & $32 \times 26$        & $4.5\pm1$       & Diffuse \\
ESO~548-G~036 & 0.20 & $16 \times 11$   & $9\pm1$    & Diffuse \\
\hline

\multicolumn{5}{p{4.4in}}{\ch Notes: Images are natural-weighted. The resolution of each
image is such that the diffuse emission is emphasized without losing much of 
the resolution.} 

\end{tabular} 
\end{center} 
\end{table}

\section{The Radio and FIR emission from the Eridanus galaxies}

\begin{figure}
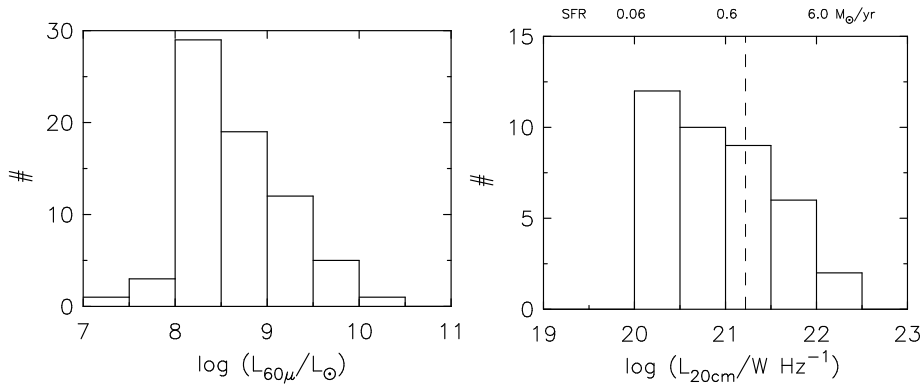

\centering

\includegraphics[width=6cm]{fir_plot1.epsi}
\includegraphics[width=6cm]{fir_plot2.epsi}

\caption{Histograms of radio and FIR luminosities of the Eridanus galaxies. The
dashed vertical line in the right hand side panel corresponds to the star
formation rate of the Milky-way.}

\label{fig:firplot} 
\end{figure}

A total of 72 Eridanus galaxies are detected in the IRAS survey, out of which
38 galaxies were detected in the NVSS. A total of 7 early type (E+S0's) galaxies
are detected in radio, most of them being radio week ($L_{20cm} \sim
10^{20}$~W~Hz$^{-1}$). The list of Eridanus galaxies detected in the FIR and in
the radio is given in Table 2. The histograms of the radio and the FIR
luminosities of the galaxies in the Eridanus group are plotted in
Fig.~\ref{fig:firplot}. The Eridanus galaxies have their radio spectral
luminosities below $\sim10^{22}$~W~Hz$^{-1}$. The high end of the radio
luminosity of the galaxies in the Hickson compact groups (Menon 1995) and the
Ursa-Major group (Verheijen \& Sancisi 2001) is consistent with that in the 
Eridanus galaxies. Majority of the late-type galaxies in nearby clusters also
have their radio luminosities below  $\sim10^{22}$~W~Hz$^{-1}$ (Reddy \& Yun
2004). It is interesting to observe that the groups (Ursa-Major, HCG's, and
Eridanus) lack in powerful AGNs ($L_{20cm} > 10^{23}$~W~Hz$^{-1}$), which are
more commonly associated with the early-type galaxies in clusters (Reddy \& Yun
2004). It can also be noticed from  Fig.~\ref{fig:firplot} that majority
($\sim70$\%) of the Eridanus galaxies have their star formation rates below that of
the Milky Way ($\sim1$~M$_{\odot}$~yr$^{-1}$). 

\section{ The Radio-FIR correlation}

\begin{figure}
\centering
\includegraphics[width=9cm, angle=0]{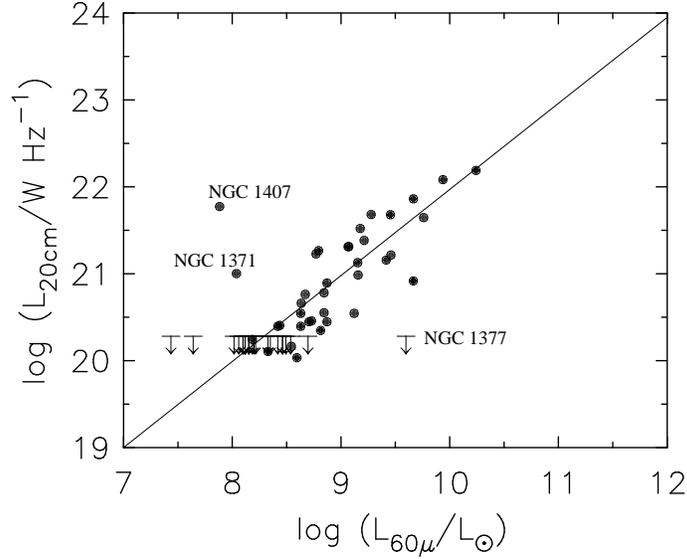}

\caption{The radio-FIR correlation for the galaxies in the Eridanus group using
the data from IRAS and NVSS. The straight line is not a fit to the data. 
It is the best fit obtained by  Yun, Reddy \& Condon (2001) for an all-sky sample of
infrared detected galaxies from IRAS.}

\label{fig:rad-fir}
\end{figure}

\begin{figure}
\centering
\includegraphics[width=9cm, angle=0]{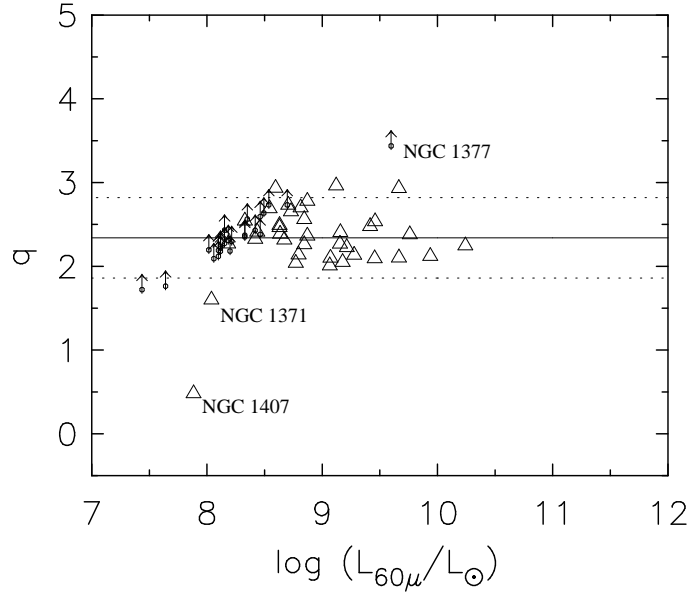}

\caption{The $q$ parameter for the Eridanus galaxies. The solid line is at
$q=2.34$ which is the mean value obtained by Yun, Reddy \& Condon (2001), and is not a
fit for the present data. The top and the bottom dotted lines are limits for
three times FIR excess and three times radio excess from the mean
respectively.}

\label{fig:q}
\end{figure}

The radio-FIR correlation for the Eridanus galaxies is shown in
Fig.~\ref{fig:rad-fir}. The $5\sigma$ upper limits (log $(L_{20cm}/W~Hz^{-1})
\sim 20.2$) for the non-detections in radio are marked by arrows. The sample
size for the Eridanus galaxies is relatively small, and does not span a large
range of infrared luminosity. Therefore, any fit is of little statistical
importance. The best fit straight line for the radio-FIR correlation obtained by
Yun, Reddy \& Condon (2001) on a large sample of IRAS galaxies is shown in 
Fig.~\ref{fig:rad-fir}. This straight line corresponds to a slope of 1 and an
intercept of 12.07. It can be seen that majority of the Eridanus
galaxies follow the radio-FIR correlation.

Following Condon, Anderson \& Helou (1991), the $q$ parameter is plotted for the Eridanus
galaxies in Fig~\ref{fig:q}. The $q$ parameter is estimated using the following
relation:

\begin{equation}
q = log \left (\frac{2.58 S_{{60\mu}} + S_{{100\mu}}}{2.98~\mathrm Jy} \right) - log
\left(\frac{S_{20cm}}{\mathrm Jy} \right)
\label{equ:qpar}
\end{equation}

In Fig.~\ref{fig:q}, the two dotted lines mark three times radio excess (below
the mean) and three time FIR excess (above the mean) respectively. 
The mean value shown by the solid
line is at $q=2.34$ as obtained by Yun, Reddy \& Condon (2001).  

It can be noticed from Figs.~\ref{fig:rad-fir} and \ref{fig:q} that there are
some galaxies which deviate significantly (more than 3 times)  from the mean
radio-FIR correlation. Rest of the galaxies follow the radio-FIR correlation
within an rms of 0.47~dex (Fig.~\ref{fig:rad-fir}).  A significant deviation in
the $q$ value can be noticed at FIR luminosities below $10^{9}$~L$_{\odot}$.
Such trends are known from earlier studies (e.g., Condon, Anderson \& Helou 1991, Yun et
al. 2001). It is believed that this non-linearity is due to the heating of  dust
from the old stellar population in galaxies (Condon, Anderson \& Helou 1991). The FIR
emission due to this heating is insignificant compared to that due to heating
from massive stars in normal star forming galaxies.

\section{The Radio-excess galaxies} 

\begin{figure}
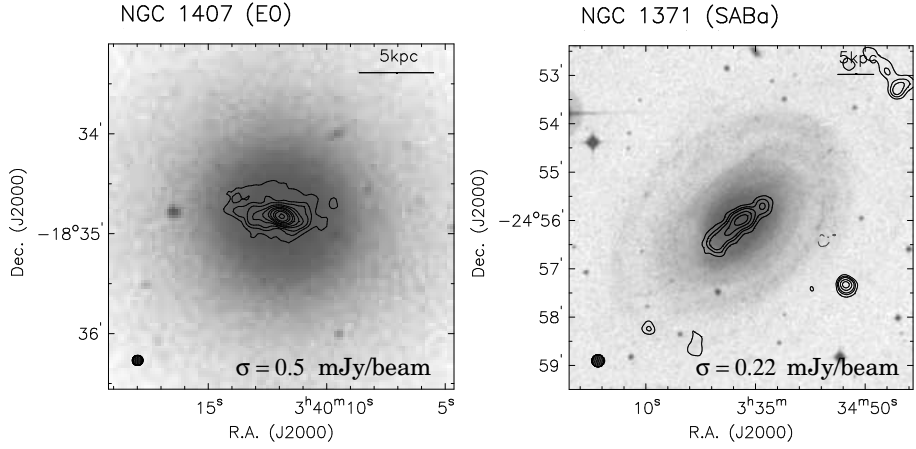

\centering
\includegraphics[width=6cm]{NGC_1407.radioC.epsi}
\includegraphics[width=6cm]{NGC_1371.radioC.epsi}

\caption{The contours of the radio continuum emission from the radio-excess
galaxies in the  Eridanus group observed with the GMRT are overlaid upon their
respective optical images from the DSS. The contours start at 3 times the rms value
($\sigma$, indicated in the map) and increases in steps of 4.5, 6, 9, 12, 18,
24, 36, 48, 72 $\times \sigma$.}

\label{fig:atlas1} 
\end{figure}

The two Eridanus galaxies, viz., NGC~1407 and NGC~1371 have significant radio
excesses. These two galaxies are marked
in Fig.~\ref{fig:q} (also, see Table 3). The radio emission
in excess of that expected from the mean radio-FIR correlation may be due to an
AGN in the galaxy. The radio continuum emission from AGNs is often from a
nuclear point source or a pair of jets or from  both. The GMRT radio continuum
contour images of these two galaxies are shown in Fig.~\ref{fig:atlas1}
overlaid upon their respective optical images from the DSS. It can be noticed that
both the galaxies have kpc-scale linear radio structures in their respective
centres. Such a feature is indicative of an AGN. A description of these two
galaxies is given below. 

\subsection{NGC~1407 (E0)}

NGC~1407 has more than 70 times excess radio emission compared to that expected
from the radio-FIR correlation. It is the most optically luminous (L$_{B} = 4
\times 10^{10}$~L$_{\odot}$) galaxy in the Eridanus group.  It resides in a
sub-group having a morphological mix of 70\% (E+S0's) and 30\% (Sp+Irr's) which
is similar to those in galaxy clusters. NGC~1407 has a nuclear X-ray source.
Diffuse X-ray emission is also seen in the Intra-group medium surrounding the
galaxy (Omar \& Dwarakanath 2004a).  This galaxy was not detected in \HI 
($M_{\HI} (5\sigma) < 1.2\times10^{7}$~M$_{\odot}$). No \HI absorption was
detected toward the radio source down to a $5\sigma$ optical depth limit of
0.05.

The GMRT radio continuum morphology (Fig.~\ref{fig:atlas1}) shows diffuse
emission along a linear structure and a nuclear continuum source coincident with
the X-ray nuclear source. The X-ray and radio nucleus, a linear radio structure,
and the large radio-excess suggest the presence of an AGN in this galaxy.

\subsection{NGC~1371 (SABa)}

NGC~1371 has more than 5 times excess radio emission compared to that expected
from the radio-FIR correlation. It is among one of the largest (D$_{B25} \sim
38$~kpc) galaxies in the Eridanus group. The optical luminosity ($L_{B} = 2
\times 10^{10}$~L$_{\odot}$) is dominated by the bulge. It has faint spiral
arms in the outer regions. The galaxy was observed in H$\alpha$ by Hameed \&
Devereux (1999). The H$\alpha$ image shows emission from HII regions in the
spiral arms, weak diffuse emission in the bulge, a nuclear source and a
ring-like structure surrounding the nucleus. The GMRT radio continuum
morphology (Fig.~\ref{fig:atlas1}) shows kpc-scale linear jet like structure
and a bright radio source coincident with the H$\alpha$ nuclear source. No
H$\alpha$ emission is seen corresponding to the extended radio emission.
Therefore, it is unlikely that the extended radio emission is due to the star
formation activities, and  could be due to a radio jet. The radio morphology,
the presence of a nuclear point source in radio and in H$\alpha$ and a
significant radio excess indicate an AGN in this galaxy. The apparent
alignment of the linear radio structure with the major axis of  NGC~1371 is
likely to be due to a projection effect although in reality the jet may be at a
considerable angle from the plane of the galaxy.

\section{The Normal star-forming galaxies}

\begin{figure}
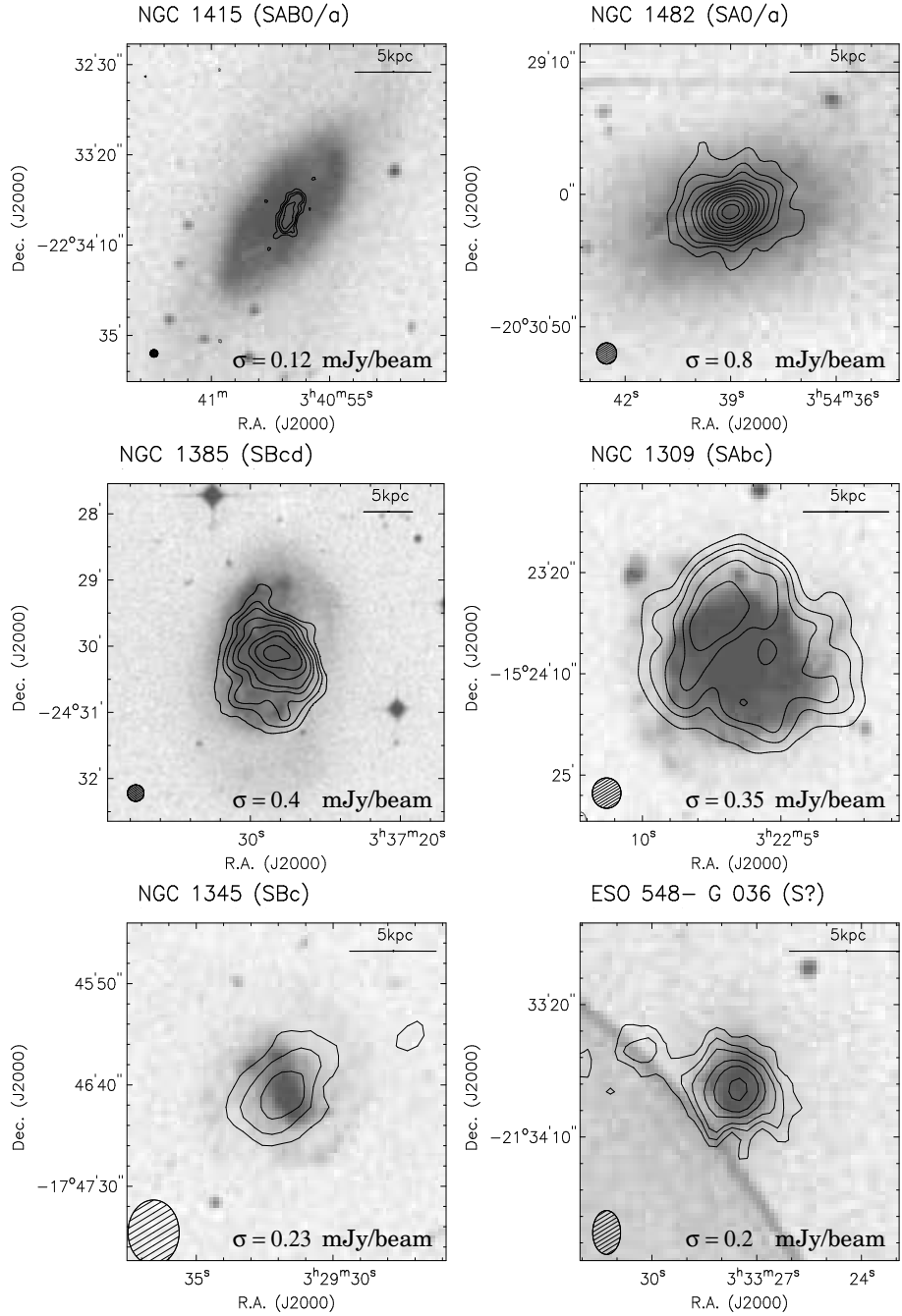

\centering

\includegraphics[width=6cm]{NGC_1415.radioC.epsi}
\includegraphics[width=6cm]{NGC_1482.radioC.epsi}
\includegraphics[width=6cm]{NGC_1385.radioC.epsi}
\includegraphics[width=6cm]{NGC_1309.radioC.epsi}
\includegraphics[width=6cm]{NGC_1345.radioC.epsi}
\includegraphics[width=6cm]{ESO_548-_G_036.radioC.epsi}

\caption{GMRT radio continuum images (in contours) of some of the normal star
forming galaxies in the  Eridanus group  overlaid upon their respective optical
images from the DSS (see Table 3). The contour levels are similar to Fig.~\ref{fig:atlas1}. The
rms values of the respective images are indicated in each image.}

\label{fig:atlas2} 
\end{figure}

The majority (70\%) of galaxies in the Eridanus group have their star formation
rates (SFR) below that of the Milky Way, viz., $\sim1$~M$_{\odot}$~yr$^{-1}$.
The SFR can be inferred from the FIR luminosity or from the radio luminosity. The
SFR ($\alpha$)  can be estimated from the FIR flux densities using the relation
(Kennicutt 1998):

\begin{equation}
\alpha = 1.7\times10^{-10} ~f \left(1 + 0.387 S_{100{\mu}m}/S_{60{\mu}m} \right) L_{60{\mu}m}
\label{equ:TFIR}
\end{equation}

The value of $f$ is quite uncertain and varies in the range $1.2-2$ for
star-burst galaxies (Sanders, Scoville \& Soifer 1991). 

The value of $\alpha$ can be estimated using the 20~cm radio spectral luminosity (Yun et
al. 2001):

\begin{equation}
\alpha \sim 6 \times 10^{-22} \frac{L_{20cm}}{W Hz^{-1}}
\label{equ:SFR}
\end{equation} 

The radio continuum images of some of the normal star forming galaxies in the
Eridanus group are shown in Fig~\ref{fig:atlas2}. The radio emission is 
diffuse in these galaxies. It can be seen that the radio continuum emission in
NGC~1309 and NGC~1385 follows the spiral arms. This is a common trend observed
in spiral galaxies where it is believed that the spiral density wave compresses
the magnetic field in the spiral arms thereby increasing the magnetic field
strength and hence the synchrotron emission.

The most FIR luminous galaxies in the group are NGC~1482 ($L_{\mathrm FIR} \sim
4.3 \times 10^{10}$~L$_{\odot}$) and NGC~1385 ($L_{\mathrm FIR} \sim 2.5 \times
10^{10}$~L$_{\odot}$).  Although these galaxies are 'star-bursts' (Table 3), they
are called 'normal' since they follow the radio-FIR correlation. 
Both of these galaxies have signatures of recent tidal
interactions in their optical images. NGC~1385 (SBcd) has a faint stellar
envelop to the south and disturbed spirals arms. NGC~1482 (SA0) has faint
stellar streamers surrounding it. The \HI tidal tails were detected in NGC~1482
(Omar \& Dwarakanath 2004b). The large amounts ($M_{H_{2}} \sim
10^{9}$~M$_{\odot}$) of molecular Hydrogen in the centres of these two galaxies
have been inferred from the CO observations (Sanders, Scoville \& Soifer 1991, Elfhag et al.
1996). Strong stellar winds indicative of a star-burst were detected in
H$_{\alpha}$ (Hameed \& Devereux 1999, Veilleux \& Rupke 2002).

\section{The FIR-excess (radio-deficient) galaxies}

Galaxies having significant excess FIR emission are not common. Yun, Reddy \& Condon
(2001) noticed 9 galaxies with $q$ value greater than 3 in a sample of nearly
1800 galaxies. A significantly higher $q$ value implies that the galaxy has
excess FIR emission than expected from the radio luminosity. Galaxies with
$L_{60\mu} \sim 10^{8}$~L$_{\odot}$ are likely to have  excess FIR emission due
to the heating of dust from the old stellar population (Condon, Anderson \& Helou 1991). These
galaxies are not considered here. The nine FIR excess galaxies of Yun, Reddy \& Condon
(2001) have $L_{60\mu} > 10^{9}$~L$_{\odot}$. The FIR excess in these galaxies
was diagnosed recently by Roussel et al. (2003) who showed that these galaxies
are most likely observed within a few Myr of the onset of an intense star
formation episode after being quiescent for at least ~100 Myr.  According to
them, this star-burst, while heating the dust, has not produced  cosmic rays
yet. Therefore, these galaxies are radio-deficient. Such galaxies must be rare
to be observed at a given epoch since the typical ages of massive stars ($\sim
1$~Myr) are only a small fraction of the ages of the star-bursts ($\sim
100$~Myr) in galaxies. The actual number of such synchrotron deficient galaxies
will depend further on the time intervals of successive star-burst events in
a galaxy.

The Eridanus group has two such galaxies, viz., NGC~1377 and IC~1953. Both
these galaxies were studied by Roussel et al. (2003). The $q$ values are
$\sim3.4$ for NGC~1377 and $\sim2.9$ for IC~1953.  These two galaxies are
marked in Fig.~\ref{fig:q}. A brief description of these two galaxies is given
below.

\subsection{NGC~1377 (S0)}

NGC~1377 is  FIR luminous (L$_{FIR} = 8 \times 10^{9}$~L$_{\odot}$) with
$S_{60\mu}/S_{100\mu} \sim 1.3$ indicating the presence of warm dust in the
galaxy. The warm IRAS galaxies are believed to be due to nuclear star-burst
probably triggered by recent interactions or mergers (Heisler \& Vader 1995). The
radio continuum emission from this galaxy is not detected in the GMRT image
down to a $5\sigma$ sensitivity limit of 1~mJy. NGC~1377 is, therefore, radio
deficient by at least 40 times.

\subsection{IC~1953 (SBd)}

\begin{figure}
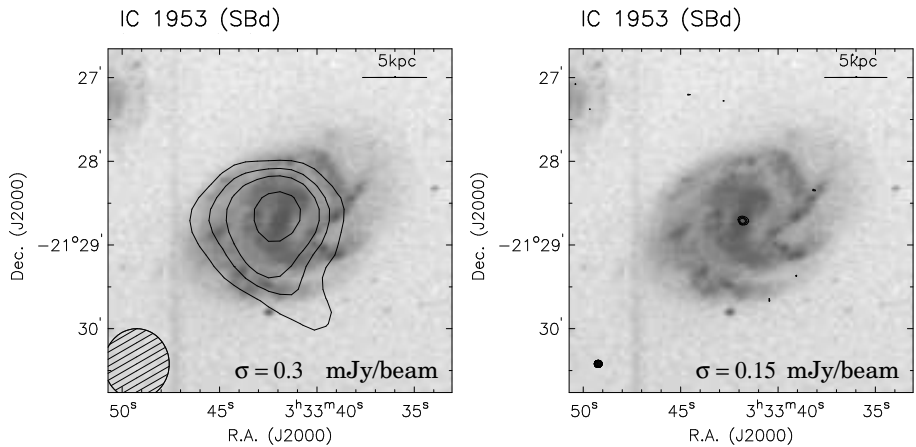

\centering
\includegraphics[width=6cm]{IC_1953.radioC.epsi}
\includegraphics[width=6cm]{IC_1953.radio2C.epsi}

\caption{The GMRT radio continuum morphologies (in contours) of IC~1953 at 
resolutions of $50''\times45''$ (left) and $5''\times5''$ (right)  overlaid on
the optical image from the DSS. The contours are at 0.9, 1.4, 1.8 and 2.7 mJy
beam$^{-1}$ for the lower resolution image and 0.5, 0.7, and 0.9 mJy
beam$^{-1}$ for the higher resolution image.}

\label{fig:atlas3}
\end{figure}

IC~1953 has grand spiral arms and a prominent bar. The radio continuum
morphology consists of diffuse emission from the disk and a nuclear point
source (Fig.~\ref{fig:atlas3}). Using high resolution (kpc-scale) mid-infrared
images from ISO (Infrared Space Observatory) observations and high resolution
radio continuum images from the VLA observations,  Roussel et al. (2003) showed
that while the disk radio continuum emission in IC~1953 is consistent with the
disk infrared emission, the nuclear radio emission is not. According to them,
it is suggestive of a recent nuclear star-burst.

\section{Conclusions}

\begin{itemize}

\item The upper end of the radio luminosity distribution of the Eridanus
galaxies ($L_{20cm} \sim 10^{20}$~W~Hz$^{-1}$) is consistent with that of field
galaxies, other groups, and late-type galaxies in clusters of galaxies.

\item The Eridanus group lacks powerful AGNs ($L_{20cm} >
10^{23}$~W~Hz$^{-1}$) commonly found in rich clusters. 

\item Most of the Eridanus galaxies follow the well-known radio-FIR correlation.

\item Majority ($\sim70$\%) of the  Eridanus galaxies have their SFR below that of
the Milky Way.

\item Two previously unknown low luminosity AGNs ($L_{20cm} <
10^{22}$~W~Hz$^{-1}$) have been detected in the group.

\item The most FIR and radio luminous galaxies in the Eridanus group have their
optical and \HI morphologies suggestive of  recent tidal interactions.

\item The Eridanus group has two FIR luminous ($L_{\mathrm FIR} \sim
10^{10}$~L$_{\odot}$) but radio-deficient galaxies.  It is believed that these
galaxies are most likely observed within a few Myr of the onset of an intense
star formation episode after being quiescent for at least ~100 Myr. 

\end{itemize}

\section*{Acknowledgments} 

We thank the staff of the GMRT who made these observations possible. The GMRT is
operated  by the National Centre for Radio Astrophysics of the Tata Institute
of Fundamental Research. This research has made use of the NASA/IPAC Infrared
archived data and Northern VLA Sky Survey data. This research has been
benefited by the NASA's Astrophysics Data System (ADS) and Extra-galactic
Database (NED) services.

\newpage
\appendix
\section{Table : FIR and radio continuum properties of the Eridanus galaxies}

The description of the table columns of Tab.~\ref{tab:sample} are as follows:

\vspace{0.1in} 

\noindent$Column~1:$ Name of the galaxy. \\ 
$Column~2:$ Hubble type. \\
$Column~3:$ $60\mu$ flux density from IRAS. \\
$Column~4:$ $100\mu$ flux density from IRAS. \\
$Column~5:$ Total FIR luminosity using Equ.~2. \\
$Column~6:$ 20~cm flux density from $NVSS$. \\
$Column~7:$ 20~cm spectral luminosity using Equ.~3. \\
$Column~8:$ Star formation rate using Equ.~6. \\

\vspace{0.1in}

\begin{table}
\begin{center}
\caption{FIR and radio properties of Eridanus galaxies}
\label{tab:sample}
\begin{tabular}{llccccrr}
\hline
\hline
\footnotesize \bf{Name} & \ch{\bf H.T.} & \footnotesize \bf{S$_{60\mu}$} & \footnotesize \bf{S$_{100\mu}$}
& \footnotesize \bf{log $\frac{L_{FIR}}{L_{\odot}}$} &\footnotesize \bf{S$_{20cm}$} & \footnotesize
\bf{log $\frac{L_{20cm}}{W Hz^{-1}}$} & \footnotesize
\bf{SFR} \\
 & &(Jy) &(Jy) & &(mJy) & &(M$_{\odot}$/yr) \\
\hline
NGC~1300 & Sbbc & 2.75 & 10.3 & 9.77 & 52.1 & 21.52 & 1.96 \\
UGCA~068 & SABm & 0.21 & 0.56 & 8.57 & -- & $<20$ & $<0.06$ \\
NGC~1325 & SAbc & 0.63 & 3.21 & 9.21 & -- & $<20$ & $<0.06$ \\
NGC~1325A & SBbc &0.24 & 0.85 & 8.70 & -- & $<20$ & $<0.06$ \\
NGC~1332 & S0 & 0.50 & 1.78 & 9.02 & 4.0 & 20.40 & 0.15 \\
NGC~1345 & SBc &0.78 & 1.59 & 9.06 & 3.9 & 20.40 & 0.15 \\
NGC~1347 & ? & 0.30 & 0.99 & 8.78 & -- & $<20$ & $<0.06$ \\
ESO~548-~G~028 & SB0 & 0.29 & 0.61 & 8.66 & -- & $<20$ & $<0.06$ \\
ESO~548-~G~029 & SB ? & 0.24 & 0.73 & 8.66 & -- & $<20$ & $<0.06$ \\
NGC~1353 & SAbc & 2.42 & 8.79 & 9.70 & 5.5 & 20.54 & 0.21  \\
IC~1952 & SBbc & 0.79 & 3.19 & 9.25 & 7.2 & 20.66 & 0.27 \\
IC~1953 & SBd & 8.47 & 11.28 & 10.05 & 13.0 & 20.92 & 0.49  \\
NGC~1359 & SBm & 2.13 & 4.28 & 9.52 & 32.2 & 21.31 & 1.21 \\
ESO~548-~G~047 & SB0&  0.39 & 1.00 & 8.83 & -- & $<20$ & $<0.06$ \\
NGC~1370 & E+ & 1.19 & 2.17 & 9.25 & 3.5 & 20.35 & 0.13 \\
NGC~1377 & S0 & 7.25 & 5.74 & 9.92 & -- & $<20$ & $<0.06$ \\
NGC~1385 & SBcd & 15.87 & 33.78 & 10.40 & 190 & 22.08 & 7.13 \\
NGC~1395 & E2 & 0.05 & 0.34 & 8.20 & -- & $<20$ & $<0.06$ \\
NGC~1400 & S0 & 0.72 & 2.50 & 9.17 & 1.7 & 20.03 & 0.06 \\
NGC~1407 & E0 & 0.14  & 0.48 & 8.45 & 93.3 & 21.77 & 3.50 \\
NGC~1415 & SAB0/a & 5.27 & 12.73 & 9.95 & 25.8 & 21.22 & 0.97 \\
ESO~482-~G~035 & SBab & 0.26 & 1.00 & 8.75 & -- & $<20$ & $<0.06$ \\
NGC~1422 & SBab & 0.39 & 1.09 & 8.85 & -- & $<20$ & $<0.06$ \\
NGC~1439 & E1 &-- & 0.34 & -- & -- & -- & $<20$ \\
NGC~1440 & SB0 & -- & 1.29 & -- & -- & -- & $<20$ \\
NGC~1438 & SB0/a & 0.23 & 0.79 & 8.67 & -- & $<20$ & $<0.06$ \\
NGC~1481 & SA0 & 0.36 & -- & 8.50 & 3.7 & 20.37 & 0.14 \\
ESO~548-~G~036 & S? & 1.94 & -- & 9.23 & 7.4 & 20.67 & 0.28 \\
ESO~548-~G~043 & Sa & 0.35 & -- & 8.49 & -- & $<20$ & $<0.06$ \\
UGCA~073 & E+ & 0.3 & -- & 8.42 & -- & $<20$ & $<0.06$ \\
NGC~1145 & Sc & 0.98 & 3.50 & 9.31 & 4.5 & 20.46 & 0.17 \\
NGC~1163 & SBbc & 0.64 & 1.71 & 9.06 & 2.3 & 20.17 & 0.09 \\
MCG~-03-08-057 & SAd & 0.57 & 2.39 & 9.12 & -- & $<20$ & $<0.06$ \\
NGC~1187 & SBc & 10.54 & 22.41 & 10.22 & 69.6 & 21.65 & 2.61 \\
NGC~1179 & SBd & 0.41 & 2.21 & 9.04 & -- & $<20$ & $<0.06$ \\
NGC~1231 & Sc & 0.23 & 0.58 & 8.60 & -- & $<20$ & $<0.06$ \\
MRK~1069 & ? & 1.28 & 1.86 & 9.24 & 9.5 & 20.78 & 0.36 \\
NGC~1232 & SABc & 3.47 & 21.7 & 10.02 & 75.5 & 21.68 & 2.83 \\
IC~1898 & SBc & 0.78 & 2.77 & 9.21 & 5.5 & 20.54 & 0.21 \\
NGC~1255 & SABbc & 2.99 & 11.36 & 9.81 & 38.0 & 21.38 & 1.43 \\
NGC~1258 & SABcd & 0.28 & 1.09 & 8.79 & -- & $<20$ & $<0.06$\\
NGC~1292 & SAc & 1.36 & 4.38 & 9.43 & 4.4 & 20.45 & 0.17 \\
UGCA~064 & SBcd & 0.19 & 0.91 & 8.68 & -- & $<20$ & $<0.06$ \\
NGC~1302 & SABa & 0.26 & 1.74 & 8.91 & -- & $<20$ & $<0.06$ \\
NGC~1306 & Sb & 2.63 & 4.73 & 9.59 & 15.2 & 20.99 & 0.57 \\
IRAS~03191-2456 &S0 & 0.53 & 0.78 & 8.86 & -- & $<20$ & $<0.06$ \\
\end{tabular} 
\end{center} 
\end{table}

\begin{table}
\begin{center}
\begin{tabular}{llccccrr}
NGC~1309 & SAbc & 5.22 & 14.26 & 9.97 & 75.2 & 21.68 & 2.82 \\
UGCA~071 & Sd & 0.28 & 0.81 & 8.71 & 2.76 & 20.24 & 0.10 \\
NGC~1338 & SABb & 1.36 & 4.95 & 9.46 & 12.3 & 20.89 & 0.46 \\
ESO~481-~G~029 & S0 & 0.48 & 1.16 & 8.91 & -- & $<20$ & $<0.06$ \\
ESO~418-~G~008 & SBd & 0.48 & 1.22 & 8.92 & 3.9 & 20.40 & 0.15 \\
NGC~1357 & SAab & 0.93 & 4.67 & 9.38 & 4.4 & 20.45 & 0.17 \\
NGC~1371 & SABa & 0.20 & 1.36 & 8.80 & 15.8 & 21.00 & 0.59 \\
NGC~1398 & SBab & 1.14 & 8.96 & 9.60 & 29.0 & 21.27 & 1.09 \\
NGC~1425 & SAb & 1.08 & 5.89 & 9.47 & 26.6 & 21.23 & 1,00 \\
NGC~1421 & SABbc & 8.48 & 21.32 & 10.16 & 114.5 & 21.86 & 4.30 \\
MCG~-02-10-009 & Sc & 0.53 & 2.13 & 9.07 & -- & $<20$ & $<0.06$ \\
MCG~-03-10-042 & SABbc & 0.86 & 3.40 & 9.28 & 9.1 & 20.76 & 0.34 \\
J034559-1231 & ? &0.63 & 1.68 & 9.05 &  & $<20$ & $<0.06$ \\
MCG~-03-10-045 & IBm & 4.77 & 7.88 & 9.83 & 22.6 & 21.16 & 0.85 \\
NGC~1461 & SA0 & 0.08 & 0.31 & 8.24 & -- & $<20$ & $<0.06$ \\
UGCA~085 & Sc & 0.24 & 0.93 & 8.72 & -- & $<20$ & $<0.06$ \\
NGC~1464 & ? & 2.61 & 4.89 & 9.59 & 21.1 & 21.13 & 0.79 \\
NGC~1482 & SA0 & 31.95 & 45.32 & 10.63 & 243 & 22.19 & 9.12 \\
IC~2007 & SBc & 1.28 & 2.77 & 9.31 & 5.6 & 20.55 & 0.21 \\
NGC~1518 & SBdm & 2.16 & 6.55 & 9.61 & 32.3 & 21.31 & 1.21 \\
NGC~1519 & SBb & 0.91 & 2.48 & 9.21 & -- & $<20$ & $<0.06$ \\
ESO~483-~G~013 & SA0 & 0.39 & 1.08 & 8.85 & 2.0 & 20.11 & 0.08 \\
J041343-1729 & ? & 0.98 & 1.77 & 9.16 &  & $<20$ & $<0.06$ \\
ESO~550-~G~008 & E? & 0.17 & -- & 8.17 &--  & $<20$ & $<0.06$ \\
UGCA~061 & SBm & 0.19 & -- & 8.22 & -- & $<20$ & $<0.06$ \\
IRAS~03007-1531 & ? & 0.36 & -- & 8.50 & -- & $<20$ & $<0.06$ \\

\hline
\end{tabular} 
\end{center} 
\end{table}

\begin{table}
\begin{center}
\caption{Galaxies studied with the GMRT}
\label{tab:radexcess}
\begin{tabular}{lcrcr}
\hline
\hline
\ch \bf{Name} & \ch \bf{Type}   & 
\ch $\bf {q}$ & \ch $\bf {S_{60\mu}/S_{100\mu}}$ &\ch \bf{Comments} \\
\hline
NGC~1407 & E   &0.45&0.29  &AGN, radio-excess\\
NGC~1371 & Sa  &1.8&0.15 & AGN, radio-excess\\
NGC~1415 & SB0/a  &2.5 &0.41 & normal\\
NGC~1482 & S0/a  &2.3 &0.70 &star-burst \\
NGC~1385 & SBcd  &2.1 &0.47 &star-burst \\
NGC~1377 & S0    &$>3.9$ &1.26 & radio-deficient \\
IC~1953  & SBcd  & 3.1 & 0.75 & radio-deficient \\
NGC~1309 & Sbc   &1.95 &0.37 & normal \\
NGC~1345 & Sbc   & 2.43 & 0.49 & normal \\
ESO~548-G~036 &S?  & 2.08 & -- & normal \\

\hline
\end{tabular} 
\end{center} 
\end{table}
\end{document}